\documentclass{agujournal2018}
\usepackage{wasysym}

\draftfalse

\journalname{JGR: Space Physics}

\begin{document}

\title{Revisited reference solar proton event of 23-Feb-1956: Assessment of the cosmogenic-isotope method sensitivity to extreme solar events}

%
%




\authors{Ilya G. Usoskin \affil{1,2},
  {Sergey A. Koldobskiy} \affil{3},
  {Gennady A. Kovaltsov} \affil{4},
 {Eugene V. Rozanov} \affil{5},
   {Timophei V. Sukhodolov} \affil{3},
     {Alexander L. Mishev} \affil{1,2},
  {Irina A. Mironova} \affil{6}
}

\affiliation{1}{Space Physics and Astronomy Research Unit, University of Oulu, Finland}
\affiliation{2}{Sodankyl\"a Geophysical Observatory, University of Oulu, Finland}
\affiliation{3}{National Research Nuclear University MEPhI, Moscow, Russia}
\affiliation{4}{Ioffe Physical-Technical Institute, 194021 St. Petersburg, Russia}
\affiliation{5}{PMOD/WRC, Davos and IAC ETH, Zurich, Switzerland}
\affiliation{6}{Department of Physics of Earth, Faculty of Physics, St. Petersburg State University, St.Petersburg, Russia}

\correspondingauthor{Ilya Usoskin}{ilya.usoskin@oulu.fi}




\begin{keypoints}
\item The integral energy spectrum of the strongest directly observed SEP event of 23-Feb-1956 is revised.
\item Sensitivity of the cosmogenic-isotope proxy method to extreme SEP events is assessed.
\item It is shown that the sensitivity of the proxy method can be significantly improved by a multi-proxy approach.
\end{keypoints}

%
%


\begin{abstract}
Our direct knowledge of solar eruptive events is limited to several decades and does not include extreme events, which can
 only be studied by the indirect proxy method over millennia, or by a large number of sun-like stars.
There is a gap, spanning 1\,--\,2 orders of magnitude, in the strength of events between directly observed and reconstructed ones.
Here, we study the proxy-method sensitivity to identify extreme solar particle events (SPEs).
First, the strongest directly observed SPE (23-Feb-1956), used as a reference for proxy-based reconstructions, was revisited
 using the newly developed method.
Next, sensitivity of the cosmogenic-isotope method to detect a reference SPE was assessed against the precision and number of
 individual isotopic records, showing that it is too weak by a factor $\approx$30 to be reliably identified in a single record.
Uncertainties of $^{10}$Be and $^{14}$C data are shown to be dominated by local/regional patterns and measurement errors, respectively.
By combining several proxy records, a SPE 4\,--\,5 times stronger than the reference one can be potentially detected, increasing the
 present-day sensitivity by an order of magnitude.
This will allow filling the observational gap in SPE strength distribution, thus enriching statistics of extreme events from 3\,--\,4
 presently known ones to several tens.
This will provide a solid basis for research in the field of extreme events, both for fundamental science, viz. solar and stellar physics,
 and practical applications, such as the risk assessments of severe space-based hazards for modern technological society.
 \end{abstract}

\section{Introduction}

Sun is an active star, which sporadically produces eruptive events, such as flares, coronal mass ejections (CMEs), interplanetary shocks, etc., in
 a wide range of energy release.
Most powerful events, particularly related to solar particle events (SPEs) with a huge enhancement of near-Earth radiation environment,
 can be hazardous for the modern technological society, especially for high- and polar low-orbit spacecraft and space missions \citep{miyake19}.
Such SPEs have been monitored since the 1940s by ground-based detectors and later by space-borne ones.
So far, 72 SPEs  have been recorded  by ground-based detectors (called ground-level enhancement, GLE --
 {https://gle.oulu.fi}), ranging from barely detectable to very strong ones.
The strongest SPE recorded so far took place on 23-Feb-1956 (GLE \#5) and had a huge enhancement of the radiation environment
 (50-fold over the galactic cosmic-ray (GCR) background, as recorded by Leeds neutron monitor) and a very hard energy spectrum of solar energetic
 particles \citep{asvestari_ASR_17}.
Quite often the strength of a SPE is quantified in terms of the event-integrated (over the entire duration $\Delta t$ of the event)
 omnidirectional fluence of particles with energy above a given threshold $E^*$:
\begin{equation}
F_{E^*} = \int_{\Delta t}\int_{E^*}^{\infty}{J(E,t)\, dE \, dt},
\end{equation}
where $J(E,t)$ is the differential energy spectrum of solar energetic particles at time $t$.
For production of cosmogenic isotopes by solar energetic particles in the Earth's atmosphere, $F_{200}$ is used
 \citep{usoskin_F200_14}, which is the event-integrated fluence with energy above 200 MeV ($E^*=200$ MeV).

\begin{figure}
\begin{center}
\includegraphics[width=1\columnwidth]{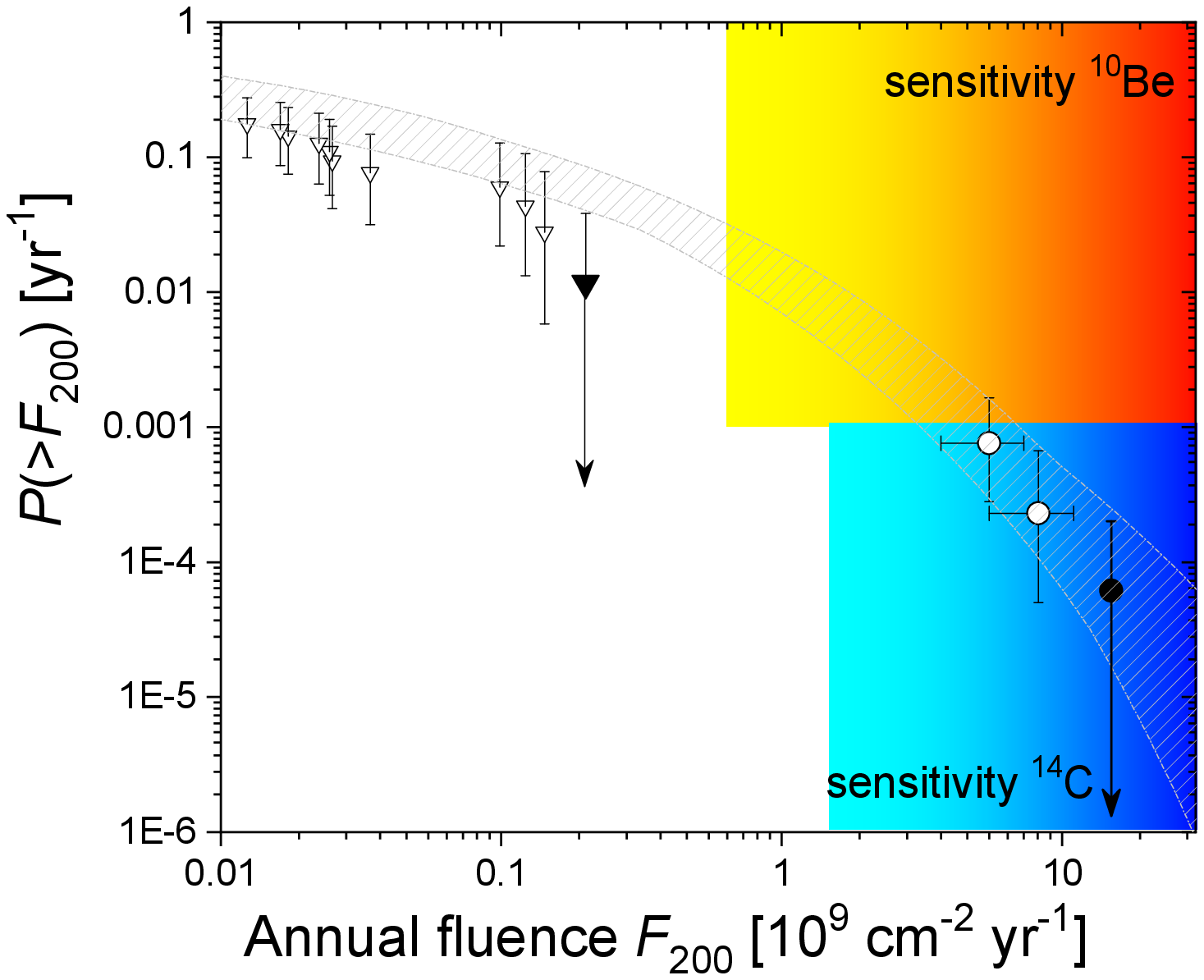}
\end{center}
\caption{Integral occurrence probability density function of the annual SPE $F_{200}$ fluence,
 according to the space era data (triangles -- data from \citet{raukunen18} and this work) and cosmogenic proxy data (circles, data from \citet{miyake19}).
The plot is modified after \citet{usoskin_F200_14}.
Open and filled symbols correspond to the known SPEs and conservative upper limits, respectively.
Error bars bound the 68\% confidence interval.
The grey-hatched area displays an estimate based on the lunar rock data \citep{poluianov18}.
The range of sensitivity of $^{14}$C- and $^{10}$Be-based reconstructions (see Section~\ref{Sec:isotopes})
 is shown by blue and yellow-red gradient filled boxes.
}
\label{Fig:F200}
\end{figure}

Statistics of the recorded event occurrence during the last decades \citep{raukunen18} are summarized in Figure~\ref{Fig:F200} as open triangles.
The plot shows the integral occurrence probability density function (IOPDF) of SPEs, which is the probability of an event with the $F_{200}$ fluence greater
 than the given value to occur within a year on average (irrespective of the solar cycle phase).
For example, the event of 23-Feb-1956 or stronger (the rightmost open triangle) has the probability to occur ranging between once per 13
 and 170 years \citep[see details in][]{usoskin_ApJ_12,usoskin_F200_14}.
The IOPDF appears quite flat, and its is hardly possible to say whether even stronger (extreme) events may occur and how often.
Straightforward extrapolations of this distribution lead to uncertainties as large as several orders of magnitude, because of the
 insufficient statistics \citep{lingenfelter80}.
Thus, until recently, estimates of the extreme SPEs parameters and occurrence probability remained highly unreliable \citep{gopalswamy18}.

The situation has changed in 2012, when an abnormal peak in carbon-14 $^{14}$C was discovered in data samples corresponding to
 the year 775 AD \citep{miyake12,usoskin_ApJ_12}.
The peak was so strong that its solar nature was initially rejected and exotic astronomical phenomena were proposed \citep{miyake12}.
However, it was soon confirmed in other cosmogenic isotope records (beryllium-10 $^{10}$Be and chlorine-36 $^{36}$Cl) and proven to be caused by a giant SPE
 \citep{usoskin_775_13,mekhaldi15,sukhodolov17}, which was a factor 50\,--\,100 stronger than the strongest directly observed SPE of 23-Feb-1956.
Although the $^{14}$C signal peaked in 775 AD, the event occurred around summer 774 AD \citep{guettler15,sukhodolov17,buentgen18,uusitalo18}.
It appeared to be a {\it black swan} -- something which is not expected to exist but it does.
It was argued that the 774 AD event was the strongest for the Sun over the last twelve millennia Holocene \citep{usoskin_LR_17},
  as depicted by the rightmost open circle in Figure~\ref{Fig:F200}.
This sets up the class of extreme solar events.
Subsequently two other, slightly weaker (60\,--\,80 \% of the one on 774 AD) but still extreme SPEs were found in cosmogenic isotope data,
 viz., 993 AD \citep{miyake13} and 660 BC \citep{park17,ohare19}.
One more event candidate was found on 3372 BC \citep{wang17} that still awaits to be confirmed.
Thus, at present we know 3\,--\,4 extreme SPEs (each a factor of 30\,--\,100 stronger than that of 23-Feb-1956) to occur during
 the last several millennia, as reflected by the circles in Figure~\ref{Fig:F200}.

The statistics based on cosmogenic data imply a roll-off of the occurrence probability of extreme SPEs.
However, there is a large gap of a factor of $\approx$30 between the space-era and the proxy-based datasets.
The gap is caused by observational techniques: on one hand, no extreme SPEs have occurred during the space era,
 and on the other hand, the sensitivity of the cosmogenic isotope method is presently insufficient to detect
 'regular' events, but only extreme SPEs.
This leaves a question open of whether the extreme SPEs form a special class of events with different distribution
 or it is a tail of the same distribution as all SPEs.
This question is crucial to study eruptive events and their terrestrial/planetary impacts, and can be answered only by
 detection of intermediate events, which are a factor 3\,--\,10 greater than that of 23-Feb-1956.
However, at present, the sensitivity of the cosmogenic isotope method to SPEs has not been studied in detail,
 the observational threshold is not defined, and the recommendations on improvements are not set.
Some earlier estimates \citep{usoskin_GRL_SCR06,mccracken15} were made in an oversimplified way, based on outdated
 cosmogenic isotope yield functions and neglected details of the atmospheric transport and deposition, which are crucial
 for reliable detection of weak signals.

Here we present the first systematic assessment of the sensitivity of the terrestrial system to record extreme SPEs
 based on up-to-date realistic isotope production and transport models and newly available high-resolution data.
First, we fix the parameters of the reference SPE of 23-Feb-1956, revisited using the most up-to-date models described in
 Section~\ref{Sec:1956}.
Then, we discuss the sensitivity of the cosmogenic isotope method to SPE detection in Section~\ref{Sec:isotopes},
 including detailed modelling of the isotope production and atmospheric transport/deposition, for both
 $^{10}$Be in ice cores and $^{14}$C in tree rings.
Finally, we set up the sensitivity of the cosmogenic-isotope method and provide recommendations on
 filling the observational gap, as well as discuss further prospectives.

\section{Reference event of 23-Feb-1956 (GLE \#5) revisited}
\label{Sec:1956}

Extreme SPEs were recorded in terrestrial cosmogenic isotope archives, which provide energy- and time-integrated responses
 rather than direct measurements of the particles spectra.
However, for modelling of the events and their impacts, energy spectra need to be determined or assumed.
It has been shown, using an analysis of strong space-era SPEs, that an event's strength correlates with the spectrum
 hardness, viz. the stronger SPE is, the harder its spectrum tends to be \citep{asvestari_ASR_17}.
This is related either to the efficiency of particle acceleration in extreme conditions, which require simultaneous appearance
 of several favorable factors, or to interplanetary transport effects, such as the so-called ``streaming limit'' \citep{reames10}.
In particular, the spectrum of the strongest directly observed SPE of 23-Feb-1956 was one of the hardest for the
 directly recorded events \citep[e.g.,][]{vashenyuk08,tuohino18}.

Estimates of the extreme SPEs, made using a multi-isotope approach, imply that their spectra were also hard and consistent
 with that of 23-Feb-1956 in the energy range below a few hundred MeV, and scaled up by different factors \citep{mekhaldi15,ohare19}.
Accordingly, the SPE of 23-Feb-1956 is typically used as a reference SPE so that it is simply scaled up to match indirectly
 found extreme SPEs \citep{usoskin_775_13,mekhaldi15,sukhodolov17}.
Here we also consider this SPE as the reference one, denoting it further as SPE$_{1956}$.
Until recently, the SPE$_{1956}$ spectrum was estimated to be in the lower energy range from ionospheric data \citep{webber07} and in higher
 energy range from neutron monitors, NMs, \citep{raukunen18}.
However, these earlier estimates of the spectrum, especially in the high-energy range, were based on an obsolete NM yield functions
 \citep{clem00} and a simplified approach.
Here we revise the spectral estimate of the SPE$_{1956}$ using a recent verified NM yield function (\citet{mishev13}
 updated as \citet{mishev20}) and an improved effective-rigidity method \citep{koldobsky_Reff_19}.

\begin{figure*}
\begin{center}
\includegraphics[width=1\textwidth]{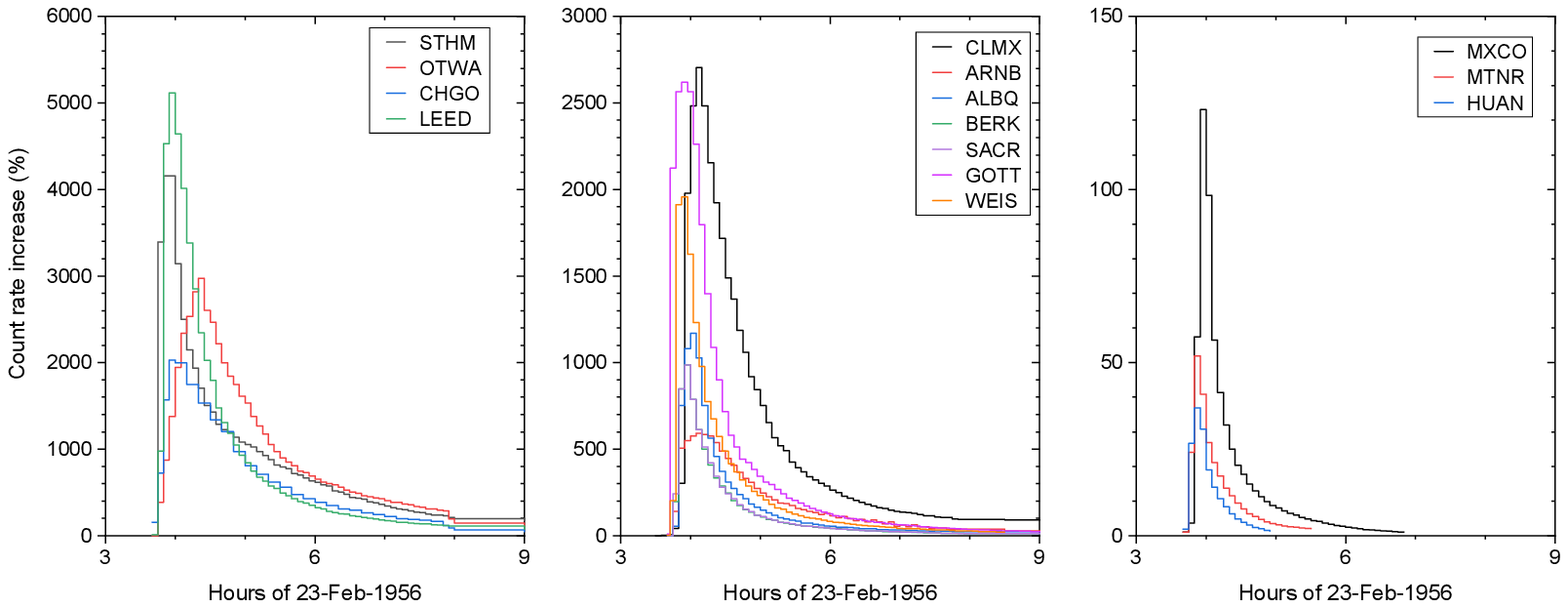}
\end{center}
\caption{Time profile of all available neutron monitors for the GLE \#5 on 23-Feb-1956 for three groups of NM locations:
 high--, mid-- and low--latitudes, from left- to right-hand side panels, respectively.
Note different Y-axes scales in the panels.
Data were obtained from the International GLE database (https://gle.oulu.fi).
}
\label{Fig:NM_data}
\end{figure*}

\begin{figure}
\begin{center}
\includegraphics[width=1\columnwidth]{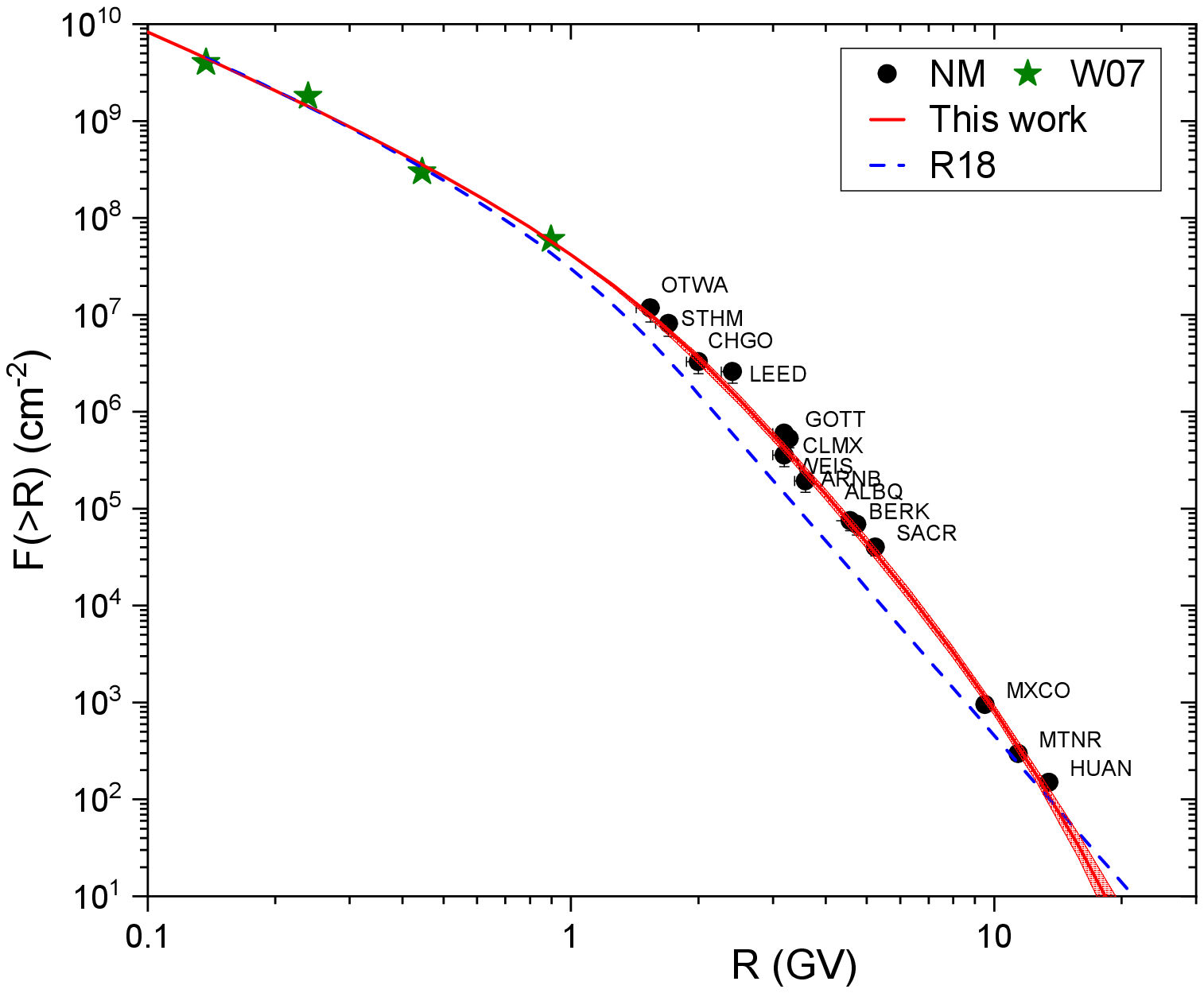}
\end{center}
\caption{Reconstructed integral rigidity spectrum of the event-integrated fluence of solar energetic particles for the SPE$_{1956}$ of
 23-Feb-1956.
Black dots are estimates based on individual neutron monitors (see Table~\ref{Tab:NM}),
 green stars represent low-energy fluence from \citet{webber07}.
The blue dashed line is the earlier spectral estimate by \citet{raukunen18}.
The red curve with the shaded area depicts the best-fit functional estimate ($J_1=1.40\cdot 10^8$ cm$^{-2}$, $\gamma_1=1.822$,
 $R_1=0.823$ GV, $J_2=1.023\cdot 10^8$ cm$^{-2}$, $\gamma_2=4.207$, $R_2=4.692$ GV and $R_b=2.380$ GV -- see Equation~\ref{Eq:spec})
 of the spectrum with the 95\% confidence interval, reconstructed here.
}
\label{Fig:Spectrum}
\end{figure}

First, we performed a full re-analysis of all the available NM data for the reference SPE, viz. GLE \#5.
The GLE was observed by 14 NMs (Table~\ref{Tab:NM}) located at different geomagnetic cutoff rigidities, from 1\,--\,13.4
 GV, as shown in Figure~\ref{Fig:NM_data}.
The highest peak count rate (5117\% above the background due to GCR) was recorded by Leeds NM, while the
 greatest event-integrated count-rate increase \citep[see definition in][]{asvestari_ASR_17} of 5276 \%*hr was observed by
 Ottawa NM.
Event-integrated responses of other NM ranged from 14.5 \%*hr for the equatorial Huancayo NM to $>$5000 \%*hr for
 high-latitude stations (see column $I$ in Table~\ref{Tab:NM}).
Next, we applied the methodology developed by \citet{koldobsky_Reff_19}, so that the effective rigidity
 $R_{\rm eff}$ and the event-integrated fluence of SEP with rigidity above it $F(>R_{\rm eff})$ were evaluated for
  each NM (see the last two columns of Table~\ref{Tab:NM}).
These points are shown as black dots with error bars in Figure~\ref{Fig:Spectrum}.

%

%
\begin{table*}
\caption{Parameters of the neutron monitors and their responses for the GLE \#5 on 23-Feb-1956.
Columns are: 1 -- name and 2 -- standard acronym; 3 -- vertical geomagnetic rigidity cutoff for the location and date;
 4, 5 and 6 -- geographical longitude, latitude, and altitude, respectively of the NM location; 7 -- reference barometric
 pressure; 8 -- GLE integral response $I$; 9 -- effective rigidity $R_{\rm eff}$; and 10 -- the event-integrated
 fluence of SEP with rigidity above $R_{\rm eff}$.
}
\begin{center}
 \fontsize{8}{12}\selectfont
\begin{tabular}{ccrrrrrrrr}
\hline
Name & Acr & $P_{\rm C}$ & Long & Lat & Alt & $P$ & $I$      & $R_{\rm eff}$ & $F(>R_{\rm eff})$ \\
     &     &(GV)& (deg)&(deg)& (m) &(mb)&(\%*hr)& (GV)          & (cm$^{-2}$) \\
\hline
Ottawa & OTWA & 1.08 & -75.7 & 45.4 & 57 & 1008 & 5276 & 1.54$^{+0.041}_{-0.113}$ & 1.18E+7$^{+2.34E+6}_{-3.21E+6}$\\
Stockholm & STHM & 1.39 & 18.0 & 59.4 & 0 & 1013 & 5129 & 1.7$^{+0.041}_{-0.113}$ & 8.13E+6$^{+1.42E+6}_{-2.00E+6}$\\
Chicago & CHGO & 1.71 & -87.7 & 41.8 & 200 & 1000 & 3392 & 2.0$^{+0.041}_{-0.124}$ & 3.29E+6$^{+5.22E+5}_{-7.90E+5}$\\
Leeds & LEED & 2.15 & -1.6 & 53.8 & 72 & 1004 & 4449 & 2.41$^{+0.041}_{-0.144}$ & 2.61E+6$^{+4.00E+5}_{-6.22E+5}$\\
Gottingen & GOTT & 2.92 & 9.9 & 51.5 & 0 & 1013 & 2112 & 3.19$^{+0.052}_{-0.196}$ & 6.00E+5$^{+8.66E+4}_{-1.42E+5}$\\
Weissenau & WEIS & 2.92 & 10.0 & 51.0 & 0 & 1013 & 1257 & 3.19$^{+0.052}_{-0.196}$ & 3.58E+5$^{+5.17E+4}_{-8.47E+4}$\\
Climax & CLMX & 3.06 & -106.2 & 39.4 & 3400 & 672 & 2917 & 3.28$^{+0.031}_{-0.155}$ & 5.32E+5$^{+6.72E+4}_{-1.03E+5}$\\
USSA Arneb & ARNB & 3.32 & 174.8 & -41.3 & 0 & 1013 & 922 & 3.58$^{+0.051}_{-0.206}$ & 1.94E+5$^{+2.78E+4}_{-4.44E+4}$\\
Albuquerq & ALBQ & 4.38 & -106.6 & 35.1 & 1575 & 800 & 794 & 4.57$^{+0.041}_{-0.216}$ & 7.52E+4$^{+9.63E+3}_{-1.51E+4}$\\
Berkley & BERK & 4.55 & -122.3 & 37.9 & 70 & 1013 & 658 & 4.74$^{+0.051}_{-0.247}$ & 6.88E+4$^{+9.32E+3}_{-1.48E+4}$\\
Sacramento & SACR & 5.1 & -105.8 & 32.7 & 3000 & 680 & 635 & 5.24$^{+0.041}_{-0.206}$ & 4.35E+4$^{+5.31E+3}_{-7.71E+3}$\\
Mexico & MXCO & 9.45 & -99.2 & 19.3 & 2274 & 779 & 51 & 9.51$^{+0.031}_{-0.216}$ & 9.60E+2$^{+1.04E+2}_{-1.26E+2}$\\
Mt.Norikura & MTNR & 11.35 & 137.5 & 36.1 & 2840 & 720 & 22 & 11.39$^{+0.041}_{-0.165}$ & 2.94E+2$^{+3.19E+1}_{-3.49E+1}$\\
Huancayo & HUAN & 13.44 & -75.3 & -12.0 & 3400 & 704 & 14.5 & 13.48$^{+0.031}_{-0.206}$ & 1.45E+2$^{+1.53E+1}_{-1.70E+1}$\\
\hline
\end{tabular}
\end{center}
\label{Tab:NM}
\end{table*}

The reconstructed spectrum was fitted with a prescribed spectral shape of the event-integrated rigidity fluence of SEP.
Here we used a modified Band-function, which includes also exponential roll-off,
 similar to the Ellison-Ramaty spectral shape \citep{ellison85}:
\begin{equation}
F(>R)= \left\{
\begin{array}{lr}
J_{1} \cdot R^{-\gamma_{1}}  \cdot \exp\left(-R/R_{1}\right),\hspace{1.4cm}~\mathrm{if}~ R<R_{b}
\\
J_{2} \cdot R^{-\gamma_{2}}  \cdot \exp\left(-R/R_{2}\right),\hspace{1.4cm}~\mathrm{if}~ R \geq R_{b},
\end{array}
\right.
\label{Eq:spec}
\end{equation}
where $F(>R)$ is the omnidirectional fluence of particles in units of cm$^{-2}$, $R$ is the rigidity expressed in gigavolts, parameters
 $\gamma_1$, $\gamma_2$, $R_1$, $R_2$ and $J_2$ are defined by fitting, and other parameters are calculated as
\begin{equation}
\begin{array}{ll}
R_b =& (\gamma_2-\gamma_1){R_1\cdot R_2\over (R_2-R_1)}\\
J_1 =& J_2 \cdot R_b^{\gamma_1-\gamma_2}\cdot \exp{\left(\gamma_2-\gamma_1\right)}
\end{array}
\end{equation}
This function is constructed so that it and its first derivative are continuous.
The data points were fitted with the prescribed spectrum using the following Monte-Carlo procedure.
First, the exact pair of values of $R_{\rm eff}$ and $F(>R_{\rm eff})$ was randomly taken from the values with
 uncertainties shown in Table~\ref{Tab:NM} independently for each NM.
The fit (Eq.~\ref{Eq:spec}) was obtained  by minimizing the logarithmic discrepancy, using NM data for higher-rigidity range
 $R\geq R_{\rm b}$ (viz. parameters $J_2$, $\gamma_2$ and $R_2$), while the lower-rigidity range was fitted using the data-points
  from \citet{webber07}.
For the low-energy data points, originally provided by \citet{webber07} without error bars, a constant 10\% error was artificially applied.
The formal value of $\chi^2$ was computed for the fit and saved.
Then, a new set of $R_{\rm eff}$ and $F(>R_{\rm eff})$ values was randomly obtained, and the fit repeated.
Overall, 1000 fits were performed.
The one corresponding to the minimum $\chi^2$ was considered as the best fit, but all fits lie very close to
 the best one.
The best-fit spectrum, in the form of Equation~\ref{Eq:spec}, is shown in red in Figure~\ref{Fig:Spectrum} along with its 95\% confidence
 interval.
The parameters of the best-fit spectrum are given in caption of Figure~\ref{Fig:Spectrum}.
The 95\% confidence-interval uncertainties of the fitting are within 10\%.

We used this best-fit spectrum as the reference SPE spectrum in the subsequent computations.

\section{Imprints in cosmogenic records}
\label{Sec:isotopes}

Cosmogenic isotopes (most useful being $^{14}$C, $^{10}$Be and $^{36}$Cl) are produced as a subproduct of the nucleonic-muon-electromagnetic
 cascade induced by energetic particles in the terrestrial atmosphere \citep[e.g.,][]{beer12}.
The bulk of terrestrial cosmogenic isotope is produced by GCR which are always present near Earth, but extreme SPEs with hard spectra also can
 produce a detectable amount of them \citep{usoskin_GRL_SCR06,webber07,mccracken15}.
This process is well studied and can be modelled via the yield-function of the isotope production \citep[see, e.g.,][]{kovaltsov12}.

Here we modelled production of the $^{10}$Be and $^{14}$C cosmogenic isotopes due to GCR and the reference SPE$_{1956}$ using the set of the yield functions,
 which provides a full 3D isotope production pattern, as computed recently by \citet{poluianov16}.

\subsection{$^{10}$Be in polar ice cores}

The beryllium $^{10}$Be isotope is unstable with the half-life of about $(1.387\pm0.012)\cdot 10^6$ years \citep[e.g.,][]{korschinek10}.
In the atmosphere, it is produced mainly by spallation of nitrogen and oxygen by nucleons of the cosmic-ray induced
 atmospheric cascade.
After production it becomes attached to atmospheric aerosols \citep[e.g.,][]{raisbeck81} and relatively quickly (within a few years) precipitates, mostly
 at mid latitudes but reaches also polar regions.
The exact atmospheric path of beryllium is affected by scavenging, stratosphere-troposphere exchange and inter-tropospheric mixing \citep[e.g.,][]{mchargue91}.
Deposition of beryllium to ice/snow, where it is subsequently measured, can be dry- or wet-dominated,
 and its concentration in ice may be additionally affected by post-depositional processes, depending on the snow accumulation rate.
Accordingly, the concentration of $^{10}$Be measured in a given ice core does not precisely reflect the global production rate,
 but may be influenced by the local/regional climate, especially on a short time scale \citep{usoskin_10Be_09}.
Moreover, strong volcanic eruption may affect beryllium deposition in polar areas \citep{baroni11,baroni19}.

Accordingly, we performed an analysis of $^{10}$Be series from different ice cores from both hemispheres and model them individually.

\subsubsection{Data}
\label{Sec:Data}

\begin{figure*}
\begin{center}
\includegraphics[width=1\textwidth]{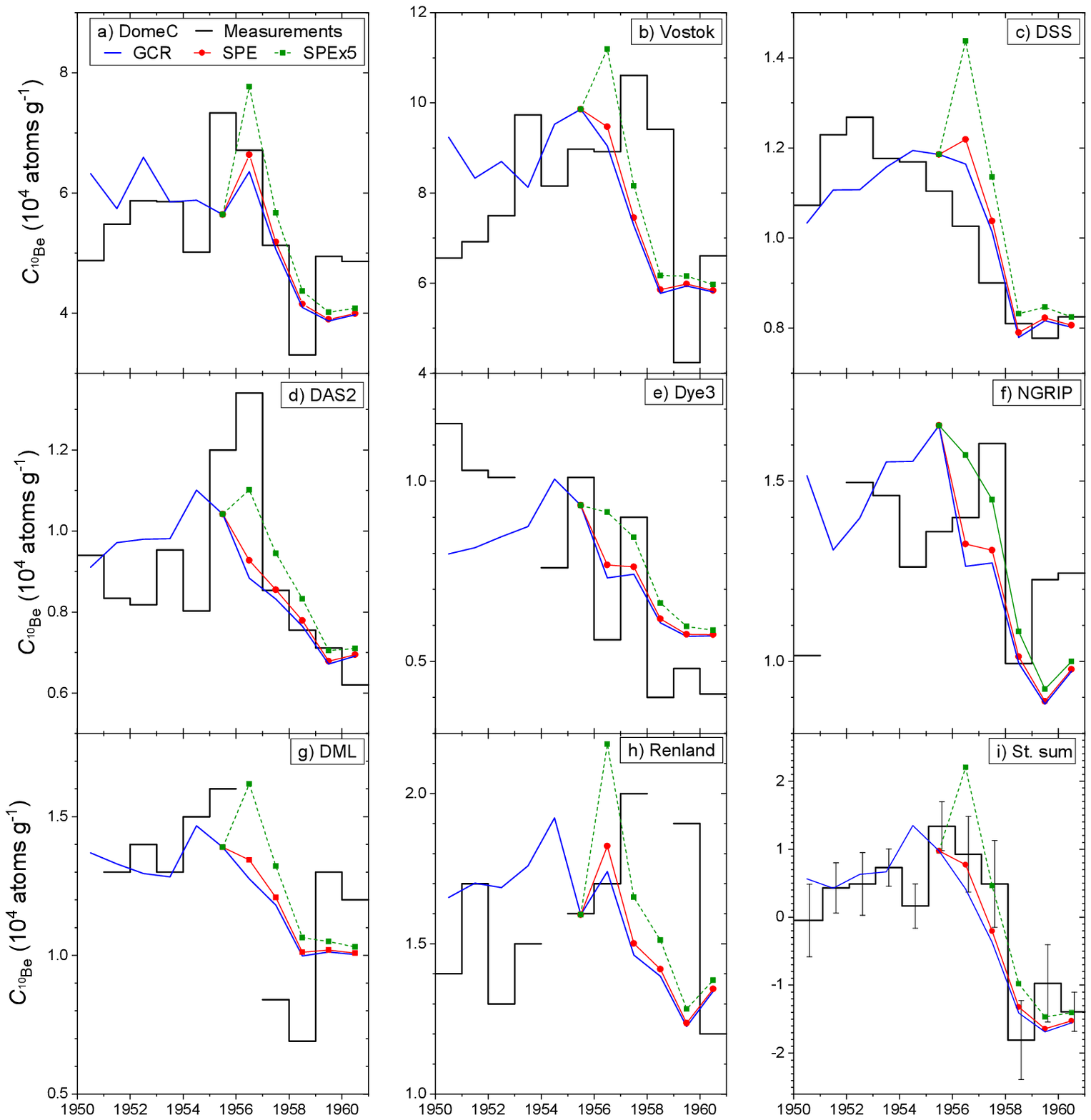}
\end{center}
\caption{Concentration of $^{10}$Be in different ice cores from Table~\ref{Tab:10Be_sites}.
Black histograms depict the measured values.
Solid blue, dotted red and dot-dashed green lines present modelled depositional fluxes of $^{10}$Be (see Section~\ref{Sec:mod-data})
 from GCR only, GCR+SPE, and GCR+SPEx5 scenarios, respectively, as denoted in the legend of panel a.
All data were reduced to the annual resolution.
Panel i depicts the standardized mean of individual series, with error bars indicating the standard error of
 the mean for each year.
}
\label{Fig:Data_annual}
\end{figure*}

We used published data of $^{10}$Be from eight ice cores or snow pits covering the decade around the SPE$_{1956}$ -- four from
 Antarctica and four from Greenland, see Table~\ref{Tab:10Be_sites}.
Since the datasets have different resolution, we reduced them all to annual $^{10}$Be concentration, as shown in Figure~\ref{Fig:Data_annual}.
While tree rings are absolutely dated within the annual accuracy, ice-core dating is less accurate, with uncertainties being relatively
 small for the last decades, viz. 1 year for high snow-accumulation sites and 1\,--\,2 years (depending to the proximity to tie points
 such as major volcanic eruptions or nuclear tests, as, e.g., the well-known peak in 1955) for low-accumulation sites \citep[see, e.g.,][]{baroni11},
 but may reach 7 years a millennium ago \citep{sigl15} or up to 70 years for the early Holocene \citep{adolphi16}.
\begin{table*}
\caption{List of polar ice cores, where $^{10}$Be measurements are available for the period 1950\,--\,1960.}
\fontsize{8}{12}\selectfont
\begin{tabular}{lcccc}
\hline
Name & location & Coordinates & Altitude & Data source\\
\hline
Concordia (Dome C) & Antarctica & 75.10$^\circ$S 123.35$^\circ$E & 3233 m & \citet{baroni19}\\
Vostok & Antarctica & 78.47$^\circ$S 106.83$^\circ$E & 3489 m & \citet{baroni11}\\
Law Dome Summit South (DSS) & Antarctica & 66.77$^\circ$S 112.8$^\circ$E & 1376 m & \citet{pedro12}\\
Dronning Maud Land (DML) & Antarctica & 71.4$^\circ$S 2.52$^\circ$W & 3200 m & \citet{aldahan98}\\
Das2 & Greenland & 67.5$^\circ$N 36.1$^\circ$W & 2936 m & \citet{pedro12}\\
Dye3 & Greenland & 65.18$^\circ$N 43.82$^\circ$W & 2480 m & \citet{beer84}\\
NGRIP & Greenland & 75.02$^\circ$N 42.53$^\circ$W & 2917 m & \citet{berggren09}\\
Renland & Greenland & 71.23$^\circ$N 26.75$^\circ$W & 2340 m & \citet{aldahan98}\\
\hline
\end{tabular}
\label{Tab:10Be_sites}
\end{table*}

\subsubsection{Modelling}
\label{Sec:10Be_mod}

We modelled production and deposition of $^{10}$Be for the period 1950\,--\,1960, covering roughly one solar cycle
 including the maximum phase of the strongest cycle \#19 (1954\,--\,1964) and the studied event.
Variability of GCR was modelled using the heliospheric modulation potential $\phi$ \citep{usoskin_Phi_05,usoskin_gil_17}
 adopted for the solar forcing incorporated in the CMIP6 (Coupled Model Intercomparison Project, stage 6) project \citep{matthes16}.
The flux of solar energetic particles during the  SPE$_{1956}$ was modelled using the fluence spectrum reconstruction described above.
All SPE-related $^{10}$Be was assumed to be produced evenly within the day of 23-Feb-1956.
3D spatial distribution (location and altitude) was modelled for each day using the input spectra of GCR and SPE$_{1956}$ by applying
 the $^{10}$Be yield function \citep{poluianov16}.

Transport and deposition of the produced $^{10}$Be were modelled using chemistry-climate model (CCM) SOCOL v.3 \citep{sukhodolov17}.
The model configuration was the same as applied for the study of 774 AD event \citep{sukhodolov17}, but all boundary conditions
 (sea surface temperature, sea ice concentration, greenhouse gas concentration, land use, geomagnetic field strength etc.) were
 prescribed for the considered period of time (1940\,--\,1960).
The model was spun-up for 10 years starting from 1940 to reach $^{10}$Be distribution similar to 1950.
GCR- and SPE-produced $^{10}$Be was traced as different species allowing us to separate their production, transport and deposition.

Depositional flux of $^{10}$Be (in units of atoms cm$^{-2}$ sec$^{-1}$) was recorded with daily resolution for the sites listed in
 Table~\ref{Tab:10Be_sites}, as shown in Figure~\ref{Fig:be10_simul} separately for the GCR- and SPE$_{1956}$-related species (blue
 and red curves, respectively).
One can see that the simulated GCR-related depositional signal is very noisy because of the regional/local weather patterns
 (curves in Figure~\ref{Fig:be10_simul} are smoothed with a 31-day running mean filter), but several features are clearly observed:
 the seasonal cycle, mostly related to the period of intrusion of stratospheric air (where a major fraction of $^{10}$Be is produced)
 into the troposphere;
 and the (inverted) solar cycle, so that almost all the blue curves have a maximum around 1954\,--\,1955 corresponding to the minimum of solar cycle \#19
 and a minimum around 1958 (maximum of the cycle \#19).
These cycles are more visible in the globally averaged $^{10}$Be deposition (Figure~\ref{Fig:be10_simul}i).
It is important to mention that the seasonal variability (roughly a factor of two) seen in the modelled $^{10}$Be deposition flux (Figure~\ref{Fig:be10_simul})
 is greater than the 11-cycle one in most of the series,
 while short sub-annual fluctuations (in particular spikes) may be also of that order of magnitude.
The modelled $^{10}$Be deposition related to the SPE$_{1956}$ is shown by red curves in Figures~\ref{Fig:Data_annual} and \ref{Fig:be10_simul},
 and it would be totally lost among the random short-term variability of the GCR-related signal and the seasonal wave pattern.
This fact makes high-resolution $^{10}$Be data not promising in recognition of SPE signals, which can be mimicked by these spikes.
A natural smoothing period for the $^{10}$Be data is annual, where the seasonal variability is averaged out.
On the other hand, the $^{10}$Be response to the SPE of 774 CE was spread over several years \citep[cf.][]{sukhodolov17} so that about half of the produced $^{10}$Be
 is deposited within the first year (viz. 1956 here), another quarter -- during the second year, and further on, so that
 each next year half of the previous is deposited, implying that the mean residence time of $^{10}$Be in the atmosphere is roughly one year \citep[e.g.,][]{raisbeck81}.

\subsubsection{Model-vs-data}
\label{Sec:mod-data}

In order to compare the model results with the measured concentration of $^{10}$Be, the annually-accumulated modelled flux was divided by the simulated
 annual water (snow) precipitation in the same grid cell.
Post-depositional effects were not taken into account.
In order to match the exact levels, the simulated values were scaled to the mean measured $^{10}$Be concentration over the period of 1950\,--\,1960.
The scaling factors appear to range from 0.8\,--\,1.6 for individual sites, which implies good agreement considering the roughness of the model spatial
 grid (2.79$^\circ \times$2.79$^\circ$) not accounting for the very local meteorology/orography, and the neglect of post-depositional effects.
The resultant modelled concentrations are shown in Figure~\ref{Fig:Data_annual} as colored curves (red and blue ones depict
 results for GCR+SPE and GCR-only related depositional flux) plotted over the measured $^{10}$Be concentrations in different ice cores
 (see Section~\ref{Sec:Data}).
One can see that the SPE$_{1956}$-related signal (the difference between the red and blue curves for the year 1956) is barely visible
 on the GCR-related background.

Although the agreement between the modelled and measured data is far from perfect, there is a clear tendency of agreement
 regarding the shape of the 11-year cycle: most of the measured data sets exhibit a decline in the last years of the series.
Some series exhibit a 1-year time mismatch between the data and model results, that may be related to the meteorological ``noise'' or
 to slightly imprecise dating.
The best agreement is observed for the DSS series.
It is clear that the SPE$_{1956}$ cannot be detected even in the simulated data, as the corresponding $^{10}$Be increase
 is only $\approx$5\% of the GCR background.
The situation is even worse for the actual noisy data with other types of possible errors not existing in the model, such as measurement
 and dating uncertainties \citep[cf.][]{mccracken15}.
Even a 5-fold SPE$_{1956}$ (SPEx5) would be not detectable over the noise in a single ice-core series (except for the DSS one) as one can see with the green
 curve in Figure~\ref{Fig:Data_annual}.
However, a 10-fold stronger event (SPEx10) would probably be detected even in a single series.
On the other hand, the standardized mean of the eight individual $^{10}$Be series (Figure~\ref{Fig:Data_annual}i)
 is much smoother than any individual one, and the SPEx5 signal stands out there at a detectable level.
Here we quantify this.

\begin{figure*}
\begin{center}
\includegraphics[width=1\textwidth]{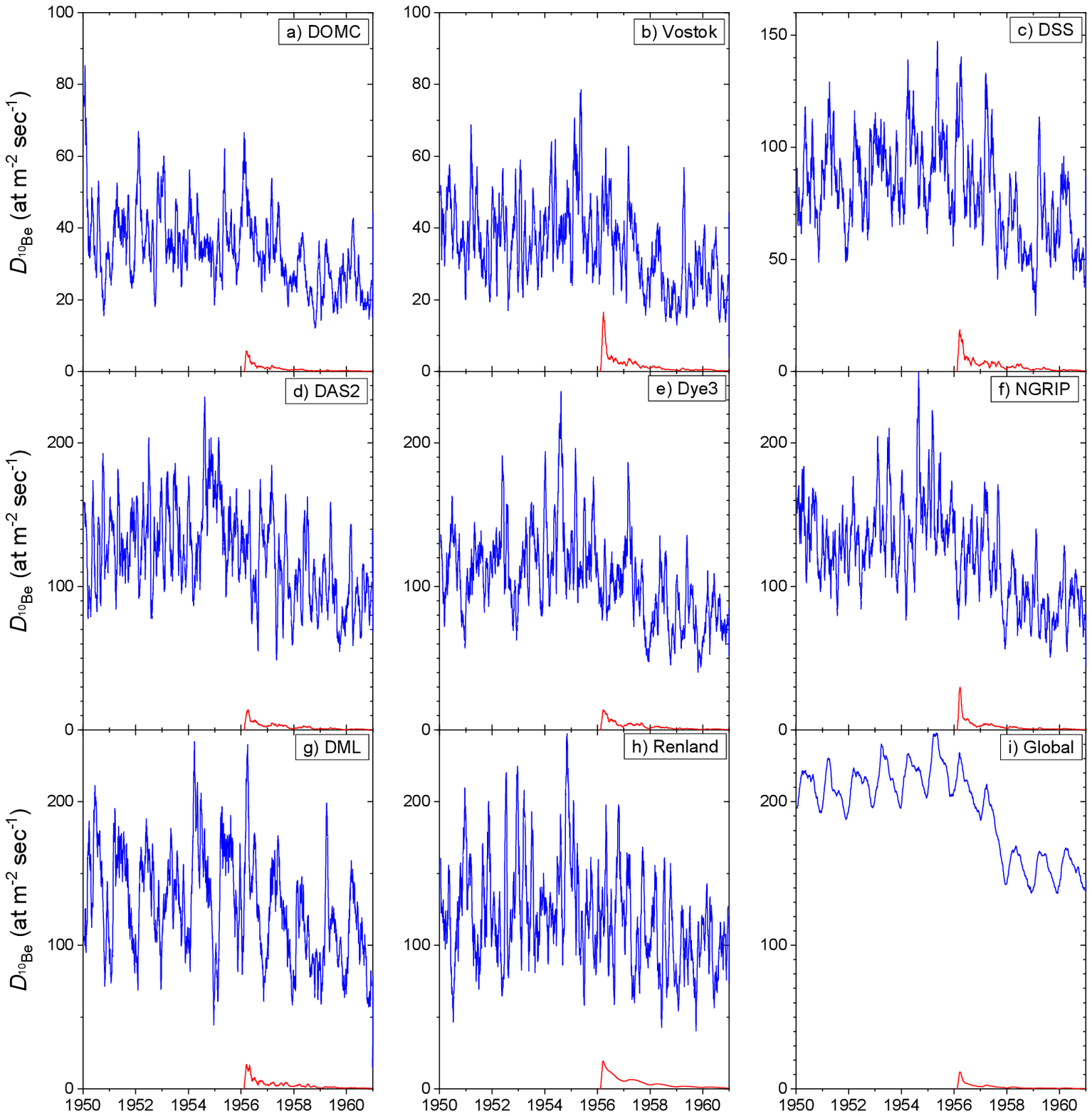}
\end{center}
\caption{Depositional flux $D_{^{10}\rm Be}$ flux of $^{10}$Be for the eight ice-core locations (see Table~\ref{Tab:10Be_sites}) and the globally averaged
 (panels {a--h} and i, respectively), simulated here using the SOCOL CCM for the period 1950\,--\,1960.
 Blue curves denote beryllium deposition due to GCR, while red ones -- contribution from the SPE$_{1956}$.
All curves are 31-day running mean smoothed.
}
\label{Fig:be10_simul}
\end{figure*}
\begin{figure}
\begin{center}
\includegraphics[width=1\columnwidth]{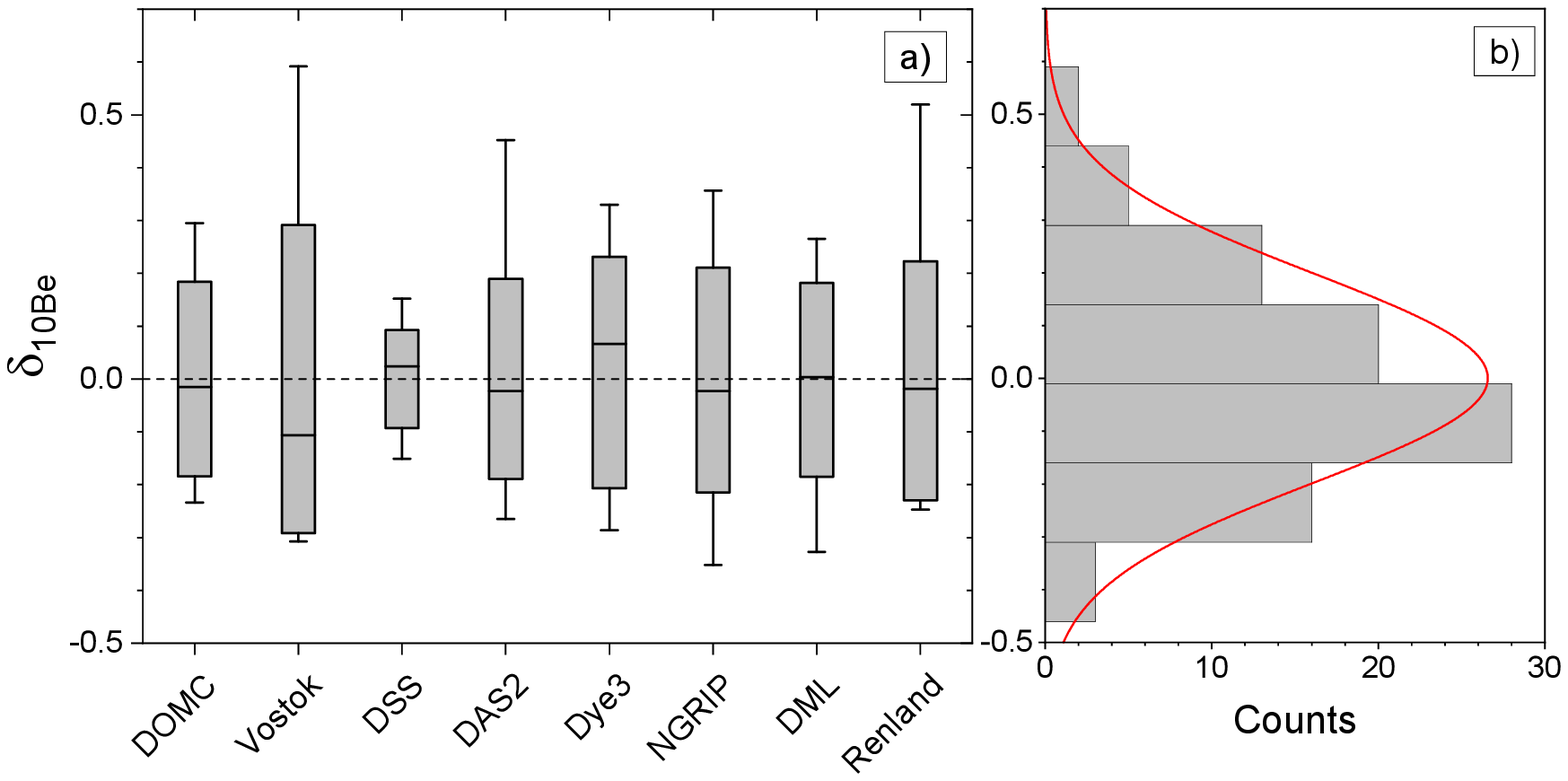}
\end{center}
\caption{Distribution of the relative differences (Equation~\ref{Eq:dBe}) between measured and
 modelled $^{10}$Be annual concentrations $\delta_{\rm 10Be}$ for the period 1950\,--\,1960.
 Panel a displays the results for individual series, where the median (horizontal lines), $\pm\sigma$ (grey boxes)
  and full range (error bars) are shown.
 Panel b displays histogram of the distribution of $\delta_{\rm 10Be}$ over all datasets,
 along with the best-fit Gaussian (zero mean, $\sigma$=0.201).
}
\label{Fig:dBe}
\end{figure}

First, for each series we have analyzed the discrepancy between the (normalized) modelled $X_{\rm mod}$ and observed $X_{\rm obs}$ data, as shown in Figure~\ref{Fig:dBe}a:
\begin{equation}
\delta_{\rm 10Be} = {X_{\rm obs}-X_{\rm mod}\over X_{\rm mod}},
\label{Eq:dBe}
\end{equation}
where $X_{\rm mod}$ is the GCR+SPE model.
For most of the series the standard deviation of $\delta_{\rm 10Be}$ is 0.2\,--\,0.3, while the DSS series depicts a significantly smaller
 discrepancy of $\approx 0.1$.
Figure~\ref{Fig:dBe}b shows the distribution of the $\delta_{\rm 10Be}$ values for all data series together.
The distribution can be well fitted with a Gaussian with zero mean and $\sigma=0.22$.
Thus, for the subsequent estimates we consider that the uncertainty of an annual $^{10}$Be data point is $\approx 20$\%.
This is significantly greater than the $1\sigma$ measurement uncertainty of $^{10}$Be concentration that is estimated as
 2\,--\,7\% \citep[e.g.,][]{pedro12, yiou97} and is largely
 dominated by the `meteorological noise', dating uncertainties or site peculiarity rather than the precision of the measurements.

\begin{figure*}
\begin{center}
\includegraphics[width=\textwidth]{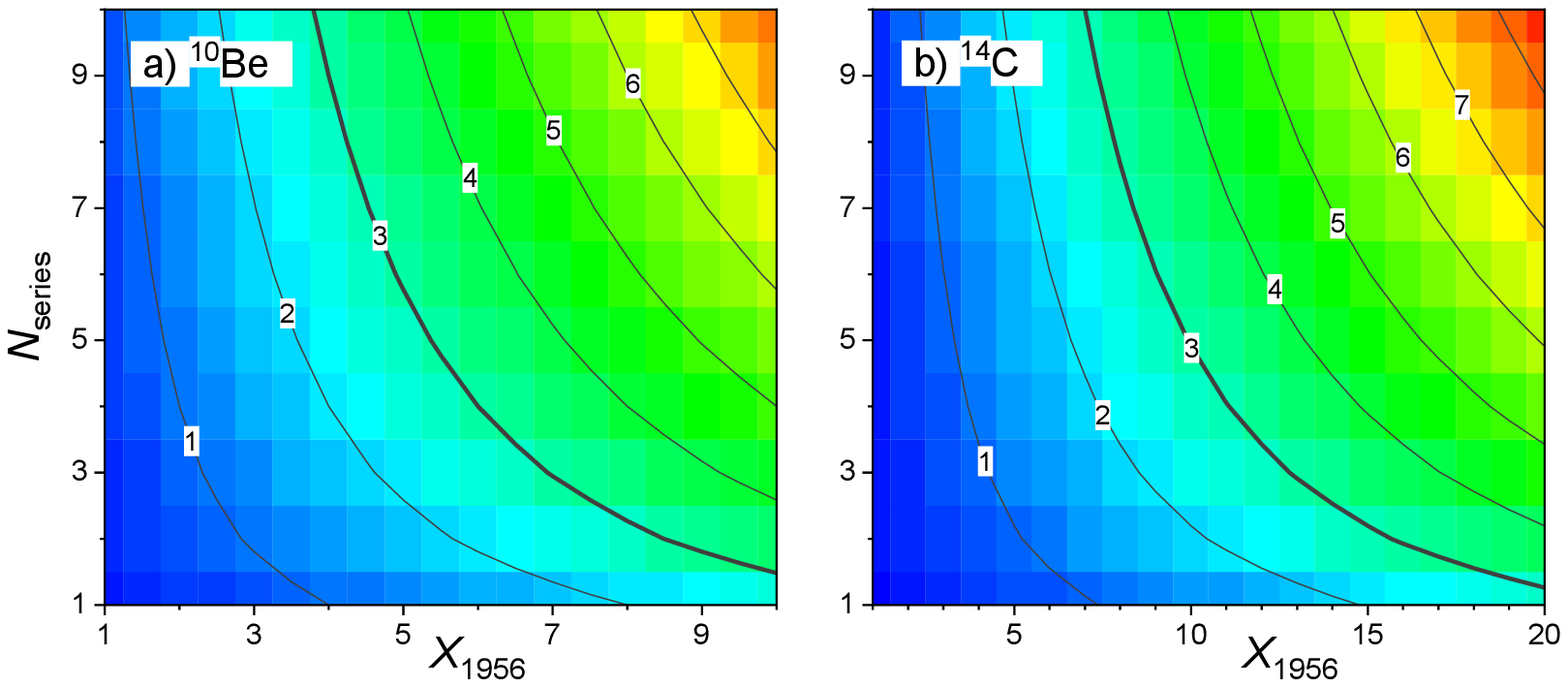}
\end{center}
\caption{a) Sensitivity of annual $^{10}$Be data to a SEP event as function of the scaled GLE \#5 ($X_{1956}$) and the number of independent data series $N_{\rm series}$.
The sensitivity (color scale) is given in units of the statistical standard deviation $\sigma$ of the measured vs. modelled $^{10}$Be signals for 1950\,--\,1960
 (as indicated by the numbers next to isolines of constant $\sigma$).
For example, an event as strong as five GLE \#5 ($X_{1956}$=5) can be identified at the $3\sigma$ level with six ice-core annual $^{10}$Be data ($N_{\rm ice}$=6).
b) The same as panel a but for $^{14}$C.
}
\label{Fig:sens}
\end{figure*}

The modelled response of polar-ice $^{10}$Be to the 2SPE$_{1956}$ is $\approx$4.8\% of the mean level (see Figure~\ref{Fig:be10_simul}),
 which is much below the uncertainty of a single annual data point of 20\%, making the signal of the reference SPE$_{1956}$
 in a single $^{10}$Be series at the $0.25\sigma$ level.
Thus, a 8x reference event would have been detected at a $2\sigma$ level, viz. significantly, even in a single $^{10}$Be series, or highly significant,
 $3.8\sigma$ if there were two independent ice cores.
The real datasets shown in Figure~\ref{Fig:Data_annual} are poorly correlated with each other so that only 2 out of 28 possible cross-correlations between
 individual series pairs appear highly significant (confidence level $>0.95$), implying that the noise is quite local.
The overall sensitivity of the $^{10}$Be data to the reference event is shown in Figure~\ref{Fig:sens}a for the scaling factor of the SPE$_{1956}$
 (X-axis) and the number of independent ice cores (Y-axis).
In a realistic case of 4\,--\,5 ice cores, a 4x reference SPE$_{1956}$ can be detected at a significant level of $2\sigma$ (0.95 confidence).
We note that increasing resolution of an isolated ice-core measurement would not enhance the sensitivity because of the seasonal cycle and
 meteorological noise in the data, but a larger number of far separated ice cores with coarser resolution could help reducing the noise.

For the existing $^{10}$Be opportunities, an optimal detection limit can be set as 4\,--\,5x reference SPE$_{1956}$.
This is illustrated in Figure~\ref{Fig:Data_annual}i, which shows the standardized mean of all the eight individual series (black histogram
 with errors bars) along with the model curve for three scenarios: only GCR (blue), GCR+SPE (red), and GCR+SPEx5 (green).
Despite a very strong diversity of the individual ice-core data series (panels a\,--\,h), the combined time profile is
 smoother and agrees well with the modelled scenario GCR+SPE: the Pearson's correlation coefficient is $r$=0.86 ($p$$\approx$6$\cdot10^{-4}$),
 $\chi^2(10)\approx17$.
We note that Antarctic sites depict a more consistent 11-year cycle pattern ($r=0.88$ between measured and modelled spatially averaged datasets)
 than Greenland-based series ($r=0.66$).

The reference SPE cannot be identified in a combined dataset of eight series, since its departure from the GCR-only scenario
 is only 0.8$\sigma$.
On the other hand, a 5-fold reference SPE (GCR+SPEx5 scenario) would have led to a $3.7\sigma$ enhancement in the year 1956,
 implying a significant identification of the peak.
This is in agreement with the modelled sensitivity shown in Figure~\ref{Fig:sens}a, where the expected detection level of a SPEx5
 event with 8 different ice-core series is $3.5\sigma$.
A SPEx10 scenario (not shown) would have been clearly identified with high statistical significance ($7.7\sigma$).

%

\subsection{Radiocarbon $^{14}$C in tree rings}

Radiocarbon $^{14}$C is produced as a result of $(n,p)$-reactions (neutron capture) by nitrogen, which makes the main sink of
 thermal neutrons in the atmosphere.
Radiocarbon gets oxidized to $^{14}$CO$_2$ and takes part in the global carbon cycles.
If absorbed by a living organism, e.g., a tree, it remains there decaying in time (half-life is about 5730 years).
The relative abundance of $^{14}$C in independently dated samples (e.g., tree rings) provides a measure of cosmic-ray flux
 at the time of the sample growing \citep[see, e.g.,][and references therein]{beer12,usoskin_LR_17}.

\begin{figure}
\begin{center}
\includegraphics[width=1\columnwidth]{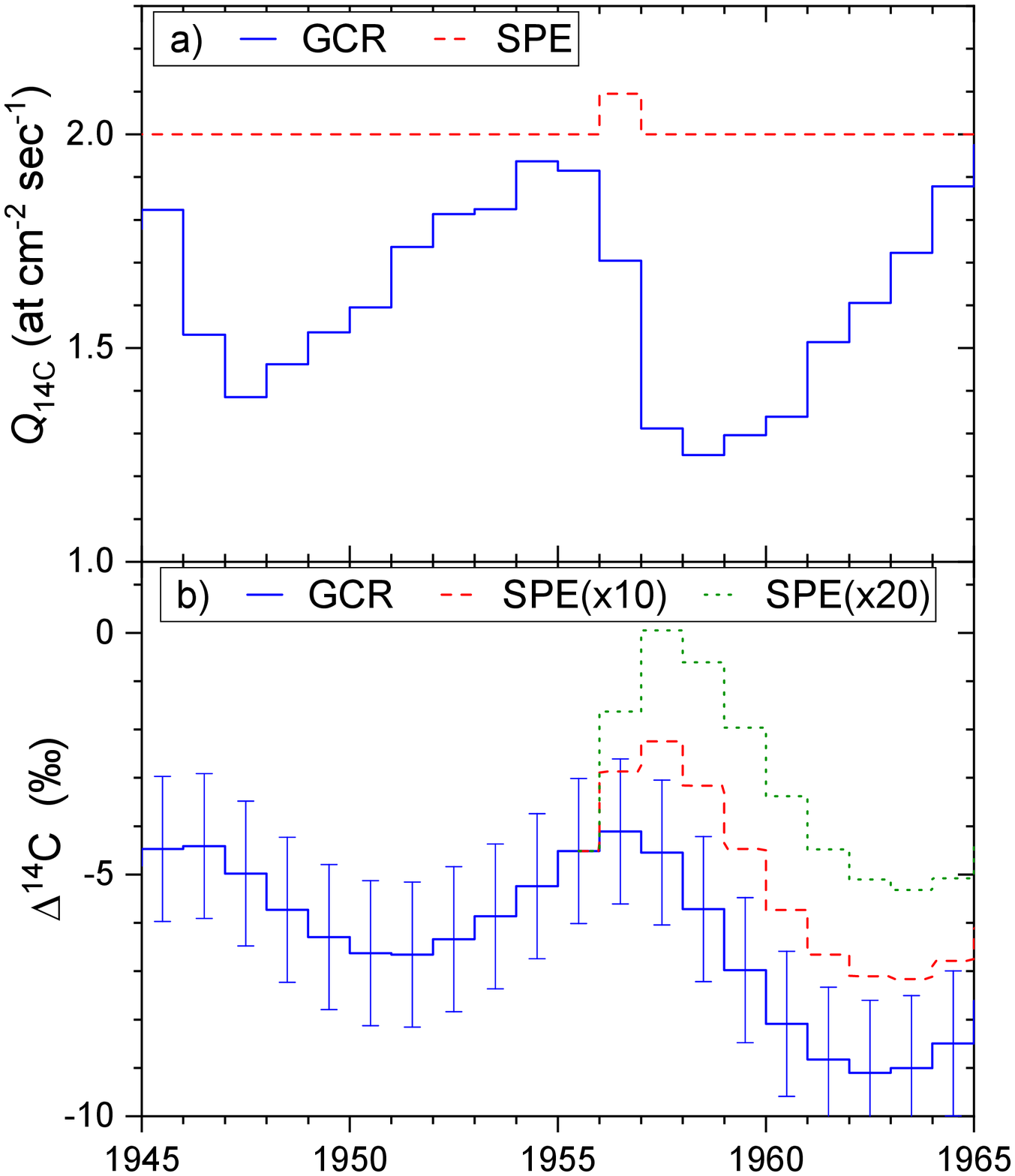}
\end{center}
\caption{Modelled annually-averaged $^{14}$C signal in the troposphere for the period 1945\,--\,1965.
Panel a) displays the modelled production rate $Q_{\rm 14C}$.
The solid blue and dashed red curves represent the GCR and reference SPE$_{1956}$ scenarios, respectively.
The SPE curve is offset by 2 (at cm$^{-2}$ sec$^{-1}$) for better visibility.
Panel b) depicts the modelled $\Delta^{14}$C for the same period.
The solid blue, dashed red and dotted green curves correspond to GCR, GCR+SPEx10 and GCR+SPEx20 scenarios, respectively.
Error bars in panel b correspond to the best modern measurement uncertainty, 1.5 \permil\, of $\Delta^{14}$C \citep{guettler15,park17}.
}
\label{Fig:C14}
\end{figure}

Unfortunately, direct $^{14}$C data cannot be used as an index of cosmic-ray intensity for the middle of the 20th century, because of the
 Suess effect (burning of fossil $^{14}$C-free fuel, which dilutes radiocarbon in the terrestrial system) and nuclear bomb tests that
 produced anthropogenic $^{14}$C well above the natural level \citep{beer12}.
On the other hand, the accuracy of modern models is sufficient to reproduce the $^{14}$C signal from known source \citep[e.g.,][]{usoskin_LR_17}.
Figure~\ref{Fig:C14} presents the modelled $^{14}$C time profile for the period under investigation.
Computations were performed using the $^{14}$C production model by \citet{poluianov16} and the 22-box carbon cycle model by \citet{buentgen18}.
The input from GCR and the SPE$_{1956}$ were treated similar to $^{10}$Be (see Section~\ref{Sec:10Be_mod}).

The modelled globally averaged $^{14}$C production rate $Q_{\rm 14C}$ is shown in Figure~\ref{Fig:C14}a as reduced
 to the annual resolution.
The (inverted) 11-year solar cycle with the $Q-$values ranging from 1.3 to 1.9 atom/cm$^2$/sec (viz. varying by a factor of $\approx$1.5)
 is clear.
Annually averaged production of $^{14}$C by the SPE$_{1956}$ is $\approx$0.1 atom/cm$^2$/sec ($3.04\cdot 10^6$ at/cm$^2$ in total),
 viz. 5\,--\,7 \% of the GCR-related level, similar to $^{10}$Be.

However, the measurable quantity, which is the relative normalized abundance of $^{14}$C with respect to $^{12}$C, considering
 isotope fractionizing, $\Delta^{14}$C \citep[e.g.,][]{beer12}, is not a direct projection of the production signal $Q$, because of the
 complicated carbon cycle which acts as a complex signal filter \citep{bard97}.
The 11-year solar cycle is shifted by 3 years ($\approx ^1$/$_4$ of the period) and attenuated by a factor of $\approx$100 to
 the magnitude of 3\,--\,5 \permil.
Since the carbon cycle effectively damps high-frequency signals, the measurable response to the SPE$_{1956}$ is attenuated even stronger,
 by a factor of $\approx 300$, down to 0.2 \permil\ (note that the red curve in Figure~\ref{Fig:C14}b is ten-fold enhanced for visibility) and spread over
 many years.
Since the modern AMS (Accelerator Mass Spectrometry) technique allows reaching the measurement uncertainties of 1.5 \permil\ in a single
 year measurement \citep[e.g.,][]{guettler15,park17}, the reference SPE$_{1956}$ yields a signal at 0.13$\sigma$ level, which is undetectable.
A 15x enhancement of the reference event is required to reach the level of significant detection in a single $^{14}$C series, which is double
 compared to $^{10}$Be.
The sensitivity of radiocarbon data to the reference event is shown in Figure~\ref{Fig:sens}b.
One can see that the sensitivity of $^{14}$C to the reference SEP event is nearly half of that for $^{10}$Be.
This is mainly due to the fact that, while beryllium, produced by a SPE, precipitates within a few years leading to a recognizable peak \citep[e.g.,][]{sukhodolov17},
 SPE-produced $^{14}$C resides in the atmosphere for many years, spreading the signal over a long period.

On the other hand, since carbon is gaseous and well mixed in the terrestrial system, uncertainties of $\Delta^{14}$C are largely defined by the
 measurement errors rather than by the regional meteorological noise.
Accordingly, almost `independent' series can be obtained by re-measuring the same samples without requiring a trees from different regions.

\section{Summary and conclusions}

Here we present a study of the sensitivity of the cosmogenic isotope method to solar energetic particles, based on the
 SPE of 23-Feb-1956, which was the strongest directly observed SPE with a very hard spectrum, and is standardly used as
 the reference SPE for analyses of the proxy-based extreme events.
First, we have revised this reference event and reconstructed the rigidity spectrum of energetic particles for it (Figure~\ref{Fig:Spectrum}),
 using the new state-of-the-art method, and have shown that earlier studies underestimated the
 strength of the event by a factor of up to three in the mid-energy range of a few GeV.
By applying this newly reconstructed spectrum, we have evaluated the expected response of the cosmogenic isotopes $^{10}$Be and
 $^{14}$C to the reference event (Figures~\ref{Fig:Data_annual}, \ref{Fig:be10_simul} and \ref{Fig:C14}).
We found out that the reference SPE cannot be reliably identified in a single proxy record \citep[cf.,][]{usoskin_GRL_SCR06,mccracken15}.
On the other hand, we have estimated the sensitivity of the proxy method to extreme SPEs, using a multi-proxy approach,
 as shown in Figure~\ref{Fig:F200}.
By combining several independent proxy records, the uncertainties can be reduced, so that a SPE by a factor 4\,--\,5 stronger
 than the reference one, viz. an order of magnitude weaker than the strongest event of 774 AD, can be potentially detected in a multi-proxy record.

Since the uncertainties of an ice-core based $^{10}$Be record are dominated by the local/regional pattern rather than by the
 measurement errors, combining results from different ice-cores
 may help significantly reducing the uncertainties.
The uncertainties of the $^{14}$C annual data are largely determined by the measurement errors, and thus, more precise and repetitive measurements
 of the same sample can improve the sensitivity to SPEs.

The ice-core based data of $^{10}$Be is estimated to have approximately double sensitivity to SPEs with respect to $^{14}$C mostly because of the strong attenuation of the
 fast SPE signal (a factor of three compared to the 11-yr solar cycle) and the spread of the SPE-produced $^{14}$C over many years.
However, this estimate does not account for possible dating errors in ice cores \citep{sigl15}, assuming that all proxy records are
 well dated and can be easily superposed.
The real situation is not as good, and the signal can be smeared.

Overall, it is possible to increase the sensitivity of the proxy method to SPEs by an order of magnitude (Figure~\ref{Fig:F200}),
 filling the observational gap between events directly observed during the space era and the extreme events discovered recently,
 and to increase statistics of extreme events during the last millennia from 3\,--\,4 known today to several tens.
This will provide a solid basis for research in the field of extreme events, both for fundamental science, viz. solar and stellar physics,
 and practical applications, such as the risk assessments of severe space-based hazards for modern technological society.

\acknowledgments
We thankfully acknowledge useful discussions with Melani\'e Baroni (Aix Marseille Universit\'e, CNRS, IRD, INRAE, Coll France,
 CEREGE, Aix-en-Provence, France).
Data of GLE recorded by NMs were obtained from the International GLE database http://gle.oulu.fi.
PIs and teams of all the ground-based neutron monitors whose data were used here, are gratefully acknowledged.
This work was partially supported by the Academy of Finland (project no. 321882 ESPERA), MEPhI Academic Excellence
 Project (contract 02.a03.21.0005), and Project RSF 20-67-46016.
The authors benefited from discussions within the ISSI International Team work (HERIOC team) and ISWAT-COSPAR S1-02 team.




\end{document}